\documentclass[sigconf]{acmart}
\usepackage{multirow}
\usepackage{makecell}
\DeclareMathOperator*{\argmin}{arg\,min}

\AtBeginDocument{}

\copyrightyear{2025}
\acmYear{2025}
\setcopyright{licensedothergov}\acmConference[KDD '25]{Proceedings of the 31st ACM SIGKDD Conference on Knowledge Discovery and Data Mining V.1}{August 3--7, 2025}{Toronto, ON, Canada}
\acmBooktitle{Proceedings of the 31st ACM SIGKDD Conference on Knowledge Discovery and Data Mining V.1 (KDD '25), August 3--7, 2025, Toronto, ON, Canada}
\acmDOI{10.1145/3690624.3709270}
\acmISBN{979-8-4007-1245-6/25/08}

\begin{document}
\title{Electron-Informed Coarse-Graining Molecular Representation Learning for Real-World Molecular Physics}

\author{Gyoung S. Na}
\authornote{Corresponding author}
\affiliation{
  \institution{Korea Research Institute of Chemical Technology}
  \city{Daejeon}
  \country{Republic of Korea}
}
\email{ngs0@krict.re.kr}

\author{Chanyoung Park}
\authornotemark[1]
\affiliation{
  \institution{Korea Advanced Institute of Science and Technology}
  \city{Daejeon}
  \country{Republic of Korea}
}
\email{cy.park@kaist.ac.kr}

\begin{abstract}
Various representation learning methods for molecular structures have been devised to accelerate data-driven chemistry. However, the representation capabilities of existing methods are essentially limited to \textit{atom}-level information, which is not sufficient to describe real-world molecular physics. Although \textit{electron}-level information can provide fundamental knowledge about chemical compounds beyond the atom-level information, obtaining the electron-level information in real-world molecules is computationally impractical and sometimes infeasible. We propose a method for learning electron-informed molecular representations without additional computation costs by transferring readily accessible electron-level information about small molecules to large molecules of our interest. The proposed method achieved state-of-the-art prediction accuracy on extensive benchmark datasets containing experimentally observed molecular physics. The source code for HEDMoL is available at \url{https://github.com/ngs00/HEDMoL}.
\end{abstract}

\ccsdesc[500]{Computing methodologies~Machine learning}
\ccsdesc[500]{Applied computing~Chemistry}
\ccsdesc[300]{Applied computing~Physics}

\keywords{Representation learning, Chemistry, Graph neural networks}

\maketitle

\section{Introduction}
Graph neural networks (GNNs) have been successfully applied to predict the physical and chemical properties of molecules by employing graph representations of molecular structures \cite{gnn_review}. Typically, a molecular structure is represented as an attributed graph $G = (\mathcal{V}, \mathcal{U}, \textbf{X}, \textbf{E})$, where $\mathcal{V}$ is a set of nodes (i.e., atoms), $\mathcal{U}$ is a set of edges (i.e., chemical bonds), $\textbf{X} \in \mathbb{R}^{|\mathcal{V}| \times d}$ is a $d$-dimensional node-feature matrix, and $\textbf{E} \in \mathbb{R}^{|\mathcal{U}| \times l}$ is an $l$-dimensional edge-feature matrix \cite{gnn_mol}. Various GNNs have been proposed to learn latent molecular embeddings based on the graph representations of the molecular structures \cite{schnet,eccnn,attfp,faenet}.

\begin{figure}
    \centering
    \includegraphics[width=0.4\textwidth]{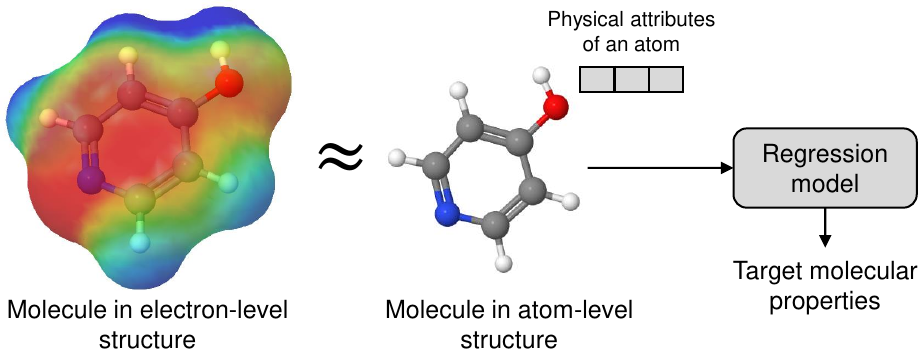}
    \caption{Basic assumptions of existing GNN-based methods in molecular representation learning, and their prediction processes on atom-level molecular structures.}
    \Description{A figure to describe the motivation of this work.}
    \label{fig:intro}
    \vspace{-4ex}
\end{figure}

Despite the numerous successes of GNN-based methods on chemical data, their representation capabilities are inherently limited in that a molecule is considered only in the \textit{atom-level} molecular structures \cite{eccnn,megnet,faenet}. Existing GNN-based methods assume that atom-level molecular structures with the physical attributes of each atom can approximate \textit{electron-level} molecular structures sufficiently, as shown in Fig. \ref{fig:intro}. However, the electronic density of molecules is inevitably distorted in the process of converting them to simplified molecular structures at the atom level \cite{dft_mol,ml_elect}. In this paper, we argue that since the physical and chemical characteristics of molecules are essentially derived from the electronic density of the molecules \cite{dft_mol}, the GNN-based methods can benefit from considering the molecular structures in the electron-level beyond the atom-level. A straightforward and direct solution would be to provide the electron-level information about the molecules to GNNs by calculating the electronic structures based on quantum mechanical calculation methods, such as density functional theory (DFT) \cite{dft_mol} and Post-Hartree–Fock methods \cite{phf_method}. However, this solution is impractical and sometimes infeasible in real-world complex molecules because the quantum mechanical calculation methods suffer from cubic or greater time complexities \cite{dft_mol} and local convergences in the electronic structure optimization \cite{dft_local_conv}.

We propose \textit{\underline{h}ierarchical \underline{e}lectron-\underline{d}erived \underline{mo}lecular \underline{l}earning} (HEDMoL) to learn electron-informed molecular representations from the input atom-level molecular structures without expensive quantum mechanical calculations. The main idea is to estimate the electron-level information about a large input molecule, which is given in an atom-level molecular structure, by extending the electron-level information about small molecules that compose the large input molecule. As the electron-level information about small molecules is readily accessible in public databases, we can relieve the computational burden of expensive quantum mechanical calculations required for large molecules. More precisely, HEDMoL learns electron-informed molecular representations from both the input atom-level molecular information and the electron-level information estimated through the following three steps: (1) HEDMoL decomposes an input atom-level molecular structure into small substructures based on graph decomposition algorithms. (2) HEDMoL constructs an electron-informed molecular graph by transferring electron-level attributes stored in a public calculation database to each of the decomposed substructures based on molecular distance, and we call this process \textit{knowledge extension}. (3) HEDMoL generates final molecular representations through hierarchical representation learning for latent embeddings of the atom-level and electron-informed molecular graphs.

It is worth noting that in this study, we focus on evaluating the prediction capabilities of the prediction models on \textit{experimental} datasets rather than \textit{simulated} datasets (e.g., QM9 \cite{qm9} and PubChemQC \cite{pubchem_qc} datasets). Although the simulated datasets are useful for analyzing rough relationships between atomic geometry and molecular properties, the simulated datasets are not appropriate to evaluate the prediction capabilities of the prediction models on real-world molecular physics because they do not sufficiently simulate the uncertainty and complex configurations in quantum mechanics \cite{acc_dft,dft_errors}. We collected eight experimentally-generated molecular datasets from physicochemistry, toxicity, and pharmacokinetics applications to evaluate the prediction capabilities of the prediction models on complex real-world molecular physics. HEDMoL achieves state-of-the-art accuracy in predicting the experimentally observed physical and chemical properties of the molecules. Furthermore, HEDMoL outperforms state-of-the-art GNNs in various regression tasks on small training datasets, which is one of the main challenges of machine learning in chemical applications \cite{tl_chem1,tl_chem2}.

\section{Related Work}
\subsection{Graph Neural Networks on Molecules}
GNNs have been widely studied to process relational data \cite{gnn_review}. Graph covolutional network (GCN) \cite{gcn} is a neural network for learning latent representations of relational data by generalizing the convolution operator. Graph attention network (GAT) \cite{gat} integrates messages from neighboring nodes based on the self-attention mechanism to learn latent node embeddings. In addition to GCN and GAT, various GNNs were proposed based on advanced message-passing schemes, such as adaptive filters \cite{egc}, transformer mechanism \cite{unimp}, and feature-wise linear modulation \cite{film}.

GNNs on 2D molecular graphs have been successfully applied to predict the physical and chemical properties of molecules \cite{gnn_mol,gnn_mol2}. A generalized framework based on message-passing neural network (MPNN) was proposed for learning quantum mechanics on atomic structures \cite{eccnn}. Directed MPNN (DMPNN) is a variant of MPNN to learn directional message propagation on 2D molecular graphs \cite{directed_mpnn}. Various GNNs were proposed for representation learning on 2D molecular graphs \cite{megnet,attfp}.

GNNs on 3D molecular graphs have also been studied in molecular science. SchNet \cite{schnet} is a graph convolutional network to learn quantum mechanics based on atom-wise representations of molecules. DimeNet \cite{dimenet} and PhysChem \cite{physchem} were devised to generate molecular embeddings by propagating the local and global atomic geometry. In addition to them, various GNNs on 3D molecular graphs have been devised \cite{m3gnet,faenet}. However, despite state-of-the-art performances of existing GNNs on several calculation datasets, their applicability in real-world molecular physics is limited because it is hard to accurately determine the 3D atomic coordinates in real-world molecules due to the uncertainty of atomic positions and the limitations of measurement equipment \cite{method_atom_coords1,method_atom_coords2}.

\subsection{Transfer Learning on Molecular Datasets}
Machine learning in chemical applications usually suffers from the lack of the training data because conducting chemical experiments to collect the training data is expensive and time-consuming \cite{cost_chem_exp1,cost_chem_exp2}. To overcome the lack of the training data, transfer learning has received significant attention in physics and chemistry \cite{tl_chem1,tl_chem2}. Several transfer learning methods have been successfully applied to various physical and chemical applications by transferring embedding networks trained on large source calculation datasets into target experimental datasets \cite{tl_chem1,tl_chem2}. Nonetheless, the prediction capability of existing transfer learning methods is inherently limited because the source calculation datasets are not able to cover the majority of large molecules in real-world experimental datasets due to the extensive time complexities of the quantum mechanical calculations for molecular structure optimization \cite{dft,dft_mol}.

In addition to conventional transfer learning methods, various molecular representation learning methods were proposed to learn informative molecular representations by transferring knowledge of decomposed substructures into the entire molecules \cite{chem_react,substructure_mol1,frag_cl}. However, the representation capabilities of the existing methods are inherently limited to atom-level information because they overlook the fundamental relationships between molecular properties and electronic densities of molecules \cite{chem_react,substructure_mol1,substructure_mol2,frag_cl}. Moreover, some existing transfer learning methods require additional information, which is not available in real-world molecules \cite{chem_react}.

\begin{figure*}
    \centering
    \includegraphics[width=0.95\textwidth]{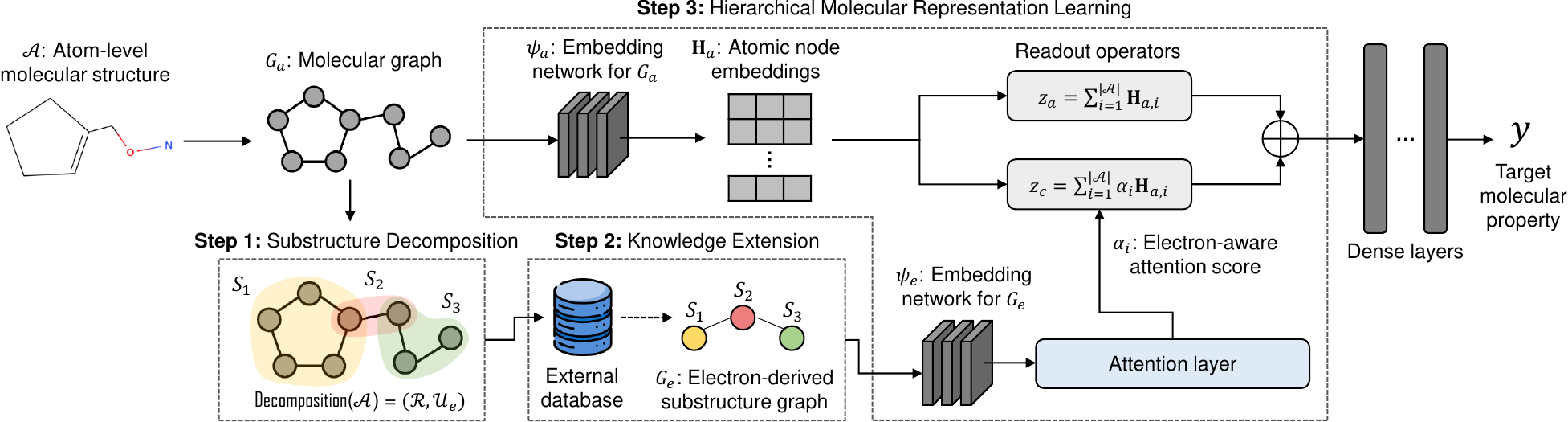}
    \caption{The overall representation learning and prediction processes of HEDMoL to predict the target molecular property $y$ of the input atom-level molecular structure $\mathcal{A}$. In the exemplary input molecule, $\mathcal{R} = \{S_1, S_2, S_3\}$ is a set of the decomposed atom-level substructures, and $\mathcal{U}_e$ is a set of edges between $\{S_1, S_2, S_3\}$.}
    \Description{Illustration of the overall architecture of the proposed method.}
    \label{fig:hedmol}
\end{figure*}

\section{Proposed Method: HEDMoL}
\subsection{Problem Reformulation}
Various methods have been proposed to learn informative molecular representation from various approaches, such as molecular geometry \cite{method_atom_coords1}, fragment-based learning \cite{substructure_mol1,frag_cl}, and domain knowledge integration \cite{chem_react}. However, their representation capabilities are fundamentally limited to the atom-level molecular structures because they overlook the principle of quantum mechanics that the physical and chemical properties of molecules fundamentally originate from their electron densities \cite{dft,dft_mol}.

Physically, a calculation process to obtain the molecular property $y$ is defined as a composite function as:
\begin{equation} \label{eq:reg_mol}
    y = (g \circ h)(\mathcal{E}),
\end{equation}
where $\mathcal{E}$ is the electronic density of a molecule, $h$ is a physics-informed function to generate a numerical representation from $\mathcal{E}$, and $g$ is a function to calculate the physical and chemical properties from the electron-level descriptors $h(\mathcal{E})$. Since the calculation methods to calculate the electronic density $\mathcal{E}$ has a cubic or greater time complexity with respect to the number of atoms, existing GNNs-based methods basically assume that $h(\mathcal{E})$ can be sufficiently approximated by the atom-level molecular structures $\mathcal{A}$ to avoid the impractical time complexity of the electronic structure calculation. However, information loss is inevitable in the approximation process $\mathcal{A} \approx h(\mathcal{E})$ \cite{dft_mol,ml_elect}.

For efficient molecular representation learning in electron-level information, we reformulate the prediction problem on the original molecules as a problem on decomposed atom-level substructures $S_1, S_2, ..., S_K$, where $K$ is the number of decomposed substructures. The reformulated prediction problem is given by:
\begin{equation} \label{eq:reform_reg_mol}
    y = (g \circ h)(\mathcal{E}_1, \mathcal{E}_2, ..., \mathcal{E}_k),
\end{equation}
where $\mathcal{E}_k$ is electronic density of the small substructure $S_k$, which is accessible in public chemical databases \cite{qm9,mp}. Based on Eq. \eqref{eq:reform_reg_mol}, we can utilize electron-level information in molecular representation learning without expensive quantum mechanical calculations by taking $\mathcal{E}_k$ from public chemical databases. The knowledge extension approach from theoretically calculated information about small atomic structures to real-world complex molecules is a long-standing challenge in computational chemistry \cite{challenges_dft}.

\subsection{Overall Architecture of HEDMoL}
Fig. \ref{fig:hedmol} illustrates the overall representation learning and prediction processes of HEDMoL, which consists of three steps as: (1) Substructure decomposition, (2) Knowledge extension, and (3) Hierarchical molecular representation learning. Each step of HEDMoL is summarized as follows. \textbf{(Step 1)} An input atom-level molecular structure $\mathcal{A}$ is decomposed into a set of atom-level substructures $\mathcal{R} = \{S_1, ..., S_K\}$. Each decomposed substructure $S_k$ implies the atom-level descriptor of $\mathcal{E}_k$. \textbf{(Step 2)} HEDMoL assigns electron-level attributes to $\mathcal{E}_k$ based on molecular distance measures on external calculation databases. The decomposed substructures are converted into an electron-derived substructure graph $G_e = (\mathcal{R}, \mathcal{U}_e, \textbf{X}_e)$, where $\mathcal{R} = \{S_1, ..., S_K\}$ is a set of nodes, $\mathcal{U}_e$ is a set of edges, and $\textbf{X}_e$ is a feature matrix containing assigned electron-level attributes. \textbf{(Step 3)} HEDMoL calculates atom- and electron-level molecular embeddings $\textbf{z}_a$ and $\textbf{z}_e$ through the GNN-based embedding networks $\psi_a$ and $\psi_e$ for the molecular graph $G_a$ and electron-derived substructure graph $G_e$, respectively. For the calculated $\textbf{z}_a$ and $\textbf{z}_e$, HEDMoL generates the final molecular embedding by concatenating the molecular embeddings $\textbf{z}_a$ and $\textbf{z}_e$, which were generated at different levels. Finally, HEDMoL predicts the target molecular property $y$ for the final molecular embedding.

\subsection{Substructure Decomposition}
The substructure decomposition of HEDMoL is a data pre-processing step to convert an input atom-level molecular structure into a tuple $(\mathcal{R}, \mathcal{U}_e)$. Physically, each decomposed substructure $S_k$ implies the atom-level representation of the decomposed electronic density $\mathcal{E}_k$ in Eq. \eqref{eq:reform_reg_mol}. In the decomposition process, we enforce a constraint $\mathcal{A} = S_1 \cup ... \cup S_K$ to prevent information loss during the substructure decomposition process. In the next knowledge extension step, the decomposed substructures will be converted into $G_e$, which contains electron-level information about $S_1, S_2, ..., S_K$.

Among various choices for the implementation of the substructure decomposition, such as spectral clustering \cite{spect_cls}, BRICS decomposition \cite{brics}, and junction tree algorithm \cite{junc_tree}, we employed the junction tree algorithm due to the following three benefits of the junction tree algorithm as: (1) The junction tree algorithm does not require a hyperparameter tuning for each input graph. (2) It provides robust graph decomposition and node clustering results for the molecular structures \cite{junc_tree}. (3) It satisfies our constraint $\mathcal{A} = S_1 \cup ... \cup S_K$ in the substructure decomposition of HEDMoL. Therefore, HEDMoL decomposes the input atom-level molecular structure $\mathcal{A}$ into a set of atom-level substructures $\mathcal{R} = \{S_1, ..., S_K\}$ based on the junction tree algorithm, where $S_k$ is defined as a vertex clique in the entire junction tree of $\mathcal{A}$.

\subsection{Knowledge Extension} \label{sec:knowl_ext}
To avoid the computational bottleneck in the electronic structure calculation, the knowledge extension step of HEDMoL aims to transfer readily accessible electron-level information from an external calculation database $\mathcal{D}_s$ to each of the decomposed substructures $S_k\in\mathcal{R}$ instead of calculating the electronic structures. That is, the decomposition result $(\mathcal{R}, \mathcal{U}_e)$ is converted into the electron-derived substructure graph $G_e$ through the knowledge extension step. In the implementation of HEDMoL, we used the QM9 dataset \cite{qm9} as an external calculation database for the knowledge extension step.

In the knowledge extension step, HEDMoL matches each $S_k$ to a small molecule in $\mathcal{D}_s$ by calculating their molecular distance. Then, HEDMoL assigns the electron-level attributes of the selected small molecule to the feature vector of $S_k$. Formally, this information retrieval process on $S_k$ and $\mathcal{D}_s$ is given by:
\begin{equation}
    \textbf{X}_{e,k} = \textbf{q}_{\texttt{idx}_k},
    \label{eq:search}
\end{equation}
where $\textbf{X}_{e,k}$ is the $k$-th row vector of $\textbf{X}_e$. $\textbf{q}_{\texttt{idx}_k}\in\mathbb{R}^q$ is the pre-calculated electron-level attributes of a small molecule in $\mathcal{D}_s$, where $\texttt{idx}_k$ is the index of the selected small molecule in $\mathcal{D}_s$, which is the most similar to $S_k$, and $\texttt{idx}_k$ is formally calculated by:
\begin{equation}
    \texttt{idx}_k = \argmin_{i \in \{1, 2, ..., |\mathcal{D}_s| \}} ||\pi(S_k) - \pi(G_{s,i})||_2,
\end{equation}
where $G_{s,i}$ is the atom-level graph of the $i$-th molecule in $\mathcal{D}_s$, $\pi$ is a distance metric between two molecular graphs. we implemented $\pi$ based on a robust and efficient unsupervised graph embedding method called geometric scattering on graphs (GeoScattering) \cite{geos}. The prediction capabilities of HEDMoL for different implementations of $\pi$ are experimentally evaluated in Appendix \ref{sec:r2_by_emb}.

After the knowledge extension step, $G_e = (\mathcal{R}, \mathcal{U}_e, \textbf{X}_e)$ of the input molecule is generated by collecting $\textbf{X}_{e,k}$ to construct $\textbf{X}_e$. As a result, HEDMoL generates $G_e$ containing the electron-level information of the decomposed substructures of the input molecule without expensive quantum mechanical calculations for optimizing electronic structures. In the next step, $G_e$ will be used for the hierarchical representation learning on the electron- and atom-levels.

\subsection{Hierarchical Representation Learning} \label{sec:hedmol_training}
The purpose of the hierarchical representation learning step in HEDMoL is to generate a latent embedding vector $\textbf{z}$ from $G_a$ and $G_e$, which represent the molecular structure in the atom- and electron-levels. The hierarchical representation learning of HEDMoL follows the physical principle in the processes of the quantum mechanical calculations to calculate the molecular properties from the fundamental electronic densities \cite{dft,dft_mol}.

For the hierarchical representation learning, HEDMoL employs GNNs to implement $\psi_a$ and $\psi_e$ for order-invariant representation learning on graph-structured data. The atom- and electron-level node embedding matrices $\textbf{H}_a$ and $\textbf{H}_e$ are calculated by $\textbf{H}_a = \psi_a(G_a)$ and $\textbf{H}_e = \psi_e(G_e)$, respectively. For the calculated $\textbf{H}_e$, an electronic state vector $\textbf{z}_e$ is calculated by $\textbf{z}_e = \sum_{i=1}^{|\mathcal{R}|} \textbf{H}_{e,i}$, where $\textbf{H}_{e,i}$ is the $i$-th row-vector of $\textbf{H}_e$. After the embedding process of $G_e$, a molecular embedding $\textbf{z}_c$ conditioned by the electron-level information is calculated by considering the atomic contributions under the given electronic state vector $\textbf{z}_e$ as:
\begin{equation}
    \textbf{z}_c = \sum_{i=1}^{|\mathcal{V}|} \frac{\exp(f_a(\textbf{H}_{a,i} \oplus \textbf{z}_e))}{\sum_{j=1}^{|\mathcal{V}|}\exp(f_a(\textbf{H}_{a,j} \oplus \textbf{z}_e))} \textbf{H}_{a,i},
\end{equation}
where $f_a$ is a trainable neural network for calculating an electron-aware attention coefficient $\alpha_i$ of the $i$-th atom under $\textbf{z}_e$. As a result, HEDMoL calculates a molecular representation $\textbf{z}_c$ conditioned by underlying electronic densities of the input molecule without expensive quantum mechanical calculations.

HEDMoL generates the molecular embedding $\textbf{z}$ of the input molecule by concatenating the graph-level embeddings from different views as $\textbf{z} = \textbf{z}_a \oplus \textbf{z}_c$, where $\textbf{z}_a = \sum_{i=1}^{|\mathcal{V}|}\textbf{H}_{a,i}$ is the graph-level embedding vector of $G_a$. Conceptually, the molecular embedding $\textbf{z}$ contains both the atom- and electron-level molecular representations. Finally, HEDMoL predicts the target molecular properties $y = f_d(\textbf{z})$ by entering $\textbf{z}$ to the trainable dense layers $f_d$.

\subsection{Energy-Based Physical Consistency Regularization in Representation Learning}\label{sec:consistency regularization}
In the molecular representation learning of HEDMoL, we enforce that the latent atom- and electron-level embeddings of the input molecule to indicate the same potential energy, which is one of the universal quantities to describe the atomic systems \cite{dft_mol,eng_mol1,pot_eng}. Physically, the atom- and electron-level descriptors of the molecule should have the same potential energy because they are essentially derived from the same electronic density. For robust molecular representation learning, we introduce two constraints in the training process of HEDMoL as follows.
\begin{equation} \label{eq:approx_energy}
    E_{p,k} + \epsilon_k = E_{a,k} = E_{e,k}, \forall k = 1, 2, ..., |\mathcal{R}|,
\end{equation}
where $E_{p,k}$ is the calculated physical energy of the small molecule that matches with the $k$-th decomposed substructure $S_k$, $\epsilon_k \sim \mathcal{N}(0, \sigma_k)$ is an independent and identically normally distributed random variable following a normal distribution $\mathcal{N}(0, \sigma_k)$, $E_{a,k}$ is the predicted energy from the node embeddings $\textbf{H}_a$ of the nodes in $S_k$, and $E_{e,k}$ is the predicted energy from the $k$-th substructure embedding $\textbf{H}_{e,k}$. In Eq. \eqref{eq:approx_energy}, $\epsilon_k$ indicates the approximation error in transferring the physical energy of the small molecule in $\mathcal{D}_s$ to the decomposed substructures in $\mathcal{R}$.

Physically, the energy of an atomic system $\mathcal{A}$ can be described by the many-body potential energies of the atoms in $\mathcal{A}$ as \cite{many_body_eng}:
\begin{align} \label{eq:many_body_e}
    E &= \sum_{i_1,i_2}^{|\mathcal{A}|} V_2(\mathcal{H}_{i_1}, \mathcal{H}_{i_2}) + \sum_{i_1, i_2, i_3}^{|\mathcal{A}|} V_3(\mathcal{H}_{i_1},\mathcal{H}_{i_2},\mathcal{H}_{i_3}) + \cdots\nonumber\\
    & + \sum_{i_1,...,i_{|\mathcal{A}|}}^{|\mathcal{A}|} V_{|\mathcal{A}|}(\mathcal{H}_{i_1},...,\mathcal{H}_{i_{|\mathcal{A}|}}),
\end{align}
where $|\mathcal{A}|$ is the number of atoms in the atomic system $\mathcal{A}$, $V_i$ is the $i$-body potential function, and $\mathcal{H}_i$ is the local environment around the $i$-th atom. However, calculating the many-body potential energy is not practical due to its infeasible computational complexity. In machine learning, various experiments demonstrated that the message-passing scheme is an efficient approach for predicting the physical properties from the physical interactions of particles \cite{gnn_many_body1,gnn_many_body2}. To overcome the infeasible computational complexity, we follow the previous studies and approximate the many-body potential energy in Eq. (\ref{eq:many_body_e}) based on a trainable message-passing function.

We use the graph self-attention mechanism \cite{gat} to calculate the many-body potential energy of $S_k$ based on interatomic attention scores in $S_k$. Formally, in the training process of HEDMoL, the many-body potential energy of $S_k$ is calculated by based on a trainable energy function $f_e$ and a message-passing function $g_e$ as:
\begin{equation} \label{eq:approx_atom_energy}
    E_{a,k} = f_e(g_e(S_k)),
\end{equation}
where an atom-level substructure embedding $g_e(S_k)$ is given by
\begin{equation}
    g_e(S_k) = \frac{1}{|S_k|}\sum_{i\in S_k}\left(\textbf{W}\textbf{H}_{a,i} + \sum_{j\in\mathcal{N}_i \cap S_k} \beta_{i,j}\textbf{V}\textbf{H}_{a,j} \right),
\end{equation}
$|S_k|$ is the number of atoms in the substructure $S_k$, $\textbf{W}$ and $\textbf{V}$ are trainable weight matrices of $g_e$, $\mathcal{N}_i$ is a set of indices of the atoms connected to the $i$-th atom, and $\beta_{i,j}$ is an attention score between the $i$-th and $j$-th atoms. Similarly, the many-body potential energy $E_{e,k}$ of $S_k$, which is retrieved from the electron-level information, is calculated by the trainable energy function $f_e$ as:
\begin{equation} \label{eq:approx_substruct_energy}
    E_{e,k} = f_e(\textbf{H}_{e,k})
\end{equation}

Based on Eqs. \eqref{eq:approx_atom_energy} and \eqref{eq:approx_substruct_energy}, for the robust molecular representation learning of HEDMoL, we define two regularization terms $\Omega_{a,n}$ and $\Omega_{e,n}$ for the $n$-th molecule in the training dataset as:
\begin{equation} \label{eq:pcl_a}
    \Omega_{a,n} = \sum_{k=1}^{|\mathcal{R}_n|} \max\{|E_{p,k} - f_e(g_e(S_k))| - \alpha, 0\},
\end{equation}
\begin{equation} \label{eq:pcl_e}
    \Omega_{e,n} = \sum_{k=1}^{|\mathcal{R}_n|} \max\{|E_{p,k} - f_e(\textbf{H}_{e,k}))| - \alpha, 0\},
\end{equation}
where $\mathcal{R}_n$ is the set of decomposed substructures of the $n$-th molecule, $\alpha \geq 0$ is a hyperparameter to allow uncontrollable energy differences incurred by the structural differences between the decomposed substructures and the matched small molecules in the external calculation database. The hyperparameter $\alpha$ is a physically bounded variable \cite{dft_errors}, and the energy differences between small organic molecules are usually in a range from 0.1 to 0.3 electronvolts \cite{dft,dft_errors}. The hyperparameter analysis for different values of $\alpha$ is conducted in Section \ref{sec:hyperparam_analysis}. Finally, we optimize the model parameters of HEDMoL to minimize the following loss function $L$ as:
\begin{equation}
    L = \sum_{n=1}^{|\mathcal{D}|} L_p(y_n, f_d(\textbf{z}_n)) + \lambda\left(\sum_{n=1}^{|\mathcal{D}|}\Omega_{a,n} + \Omega_{e,n}\right),
    \label{eq:final_loss}
\end{equation}
where $L_p$ is a prediction loss, and $\lambda \geq 0$ is a hyperparameter to control the effect of the regularization terms $\Omega_{a,n}$ and $\Omega_{e,n}$ in the training process of HEDMoL. Hyperparameter analysis results for different values of $\lambda$ are also provided in Section \ref{sec:hyperparam_analysis}.

\section{Experiments}
To evaluate the prediction capabilities of HEDMoL, we compared the prediction accuracies of HEDMoL with those of the state-of-the-art methods on various benchmark molecular datasets containing experimentally observed molecular physics. In the experiments, we focused on evaluating the prediction capabilities of HEDMoL on the \textit{experimental} datasets rather than \textit{calculation} datasets, due to the following two reasons: (1) Evaluations on the calculation datasets (e.g., QM9 dataset) are not fair because HEDMoL exploits the external calculation databases in the knowledge extension step. (2) The experimental datasets containing the uncertainty of the atomic systems are closer to real-world molecular physics than the calculation datasets \cite{ebg}.

\textbf{Datasets.} We collected experimentally generated molecular datasets from public chemical databases, such as MoleculeNet \cite{molnet} and ChEMBL \cite{chembl}. To demonstrate the effectiveness of HEDMoL in broad scientific applications, we selected eight benchmark datasets from various application fields including physicochemistry, toxicity, and pharmacokinetics. in Table \ref{tb:mol_datasets} shows the statistics and target molecular properties of the benchmark molecular datasets.

\textbf{Competitor Methods.} We compared the prediction capabilities of HEDMoL with those of a tree-based ensemble method \cite{xgb} and state-of-the-art GNNs \cite{schnet,dimenetpp,physchem,m3gnet,faenet,gatv2,gin,egc,eccnn,unimp,megnet,directed_mpnn,attfp}. Although the 3D-GNNs are not applicable to the experimental molecular datasets because the 3D atomic coordinates are not available, we measured their prediction performances by generating the 3D molecular structures based on the force-field-based and semi-empirical calculations \cite{3d_mol_calc}. Note that the force-field-based and semi-empirical calculation method is efficient but not accurate. Furthermore, we generated XGB-Mor, XGB-FC, and XGB-MK that predict target molecular properties for input Morgan (Mor) \cite{ecfp}, functional-class (FC) \cite{ecfp}, and MACCS Key (MK) \cite{maccs_key} fingerprints, respectively. Appendix \ref{sec:competitor_methods} briefly describes the compared methods.

\begin{table}
\centering
\caption{Characteristics of the benchmark molecular datasets containing the atom-level molecular structures and their experimentally observed target properties.}
\vspace{-2ex}
\resizebox{0.48\textwidth}{!}{
\begin{tabular}{c|cccc}
\toprule
\makecell[c]{Application\\Category}
& Dataset
& Target Molecular Property
& \makecell{\# of\\Data}\\
\midrule

\multirow{3}{*}{Physicochemistry}
& Lipop \cite{chembl}
& Lipophilicity
& 4,200\\

& ESOL \cite{esol}
& Aqueous solubility
& 1,128\\

& ADMET \cite{molnet}
& Aqueous solubility
& 4,801\\
\midrule

\multirow{3}{*}{Toxicity}
& IGC50 \cite{igc50}
& Tetrahymena pyriformis toxicity
& 1,791\\

& LC50 \cite{igc50}
& Fathead minnow toxicity
& 822\\

& LD50 \cite{igc50}
& Oral rat toxicity
& 7,412\\
\midrule

\multirow{2}{*}{Pharmacokinetics}
& LMC-H \cite{chembl}
& Microsomal clearance in human
& 5,347\\

& LMC-R \cite{chembl}
& Microsomal clearance in rat
& 2,165\\
\bottomrule
\end{tabular}
}
\label{tb:mol_datasets}
\vspace{-2ex}
\end{table}

\begin{table*}
\centering
\caption{Measured $R^2$-scores on the benchmark molecular datasets. Input type means the required data format of the input molecules. The highest $R^2$-score for each benchmark dataset has been remarked in bold, and the standard deviation of the $R^2$-scores is presented in parentheses. N/R and N/A mean a negative $R^2$-score indicating a failure of regression and an execution failure related to out of memory of numerical stability problems, respectively.}
\vspace{-2ex}
\resizebox{0.95\textwidth}{!}{
\renewcommand{\arraystretch}{1.1}
\begin{tabular}{c|c|ccccccccc}
\hline
Input Type
& Method
& Lipop
& ESOL
& ADMET
& IGC50
& LC50
& LD50
& LMC-H
& LMC-R\\
\hline

\multirow{3}{*}{\makecell[c]{Molecular\\Fingerprint}}
& XGB-Mor \cite{ecfp}
& \makecell[c]{0.531 \footnotesize(0.024)}
& \makecell[c]{0.659 \footnotesize(0.045)}
& \makecell[c]{0.717 \footnotesize(0.021)}
& \makecell[c]{0.621 \footnotesize(0.040)}
& \makecell[c]{0.390 \footnotesize(0.133)}
& \makecell[c]{0.497 \footnotesize(0.016)}
& \makecell[c]{0.505 \footnotesize(0.018)}
& \bfseries\makecell[c]{0.617 \footnotesize(0.058)}\\

& XGB-FC \cite{ecfp}
& \makecell[c]{0.578 \footnotesize(0.018)}
& \makecell[c]{0.686 \footnotesize(0.052)}
& \makecell[c]{0.720 \footnotesize(0.009)}
& \makecell[c]{0.628 \footnotesize(0.023)}
& \makecell[c]{0.501 \footnotesize(0.052)}
& \makecell[c]{0.519 \footnotesize(0.025)}
& \makecell[c]{0.503 \footnotesize(0.007)}
& \bfseries\makecell[c]{0.612 \footnotesize(0.015)}\\

& XGB-MK \cite{maccs_key}
& \makecell[c]{0.542 \footnotesize(0.041)}
& \makecell[c]{0.764 \footnotesize(0.047)}
& \makecell[c]{0.761 \footnotesize(0.020)}
& \makecell[c]{0.680 \footnotesize(0.037)}
& \makecell[c]{0.486 \footnotesize(0.112)}
& \makecell[c]{0.526 \footnotesize(0.021)}
& \makecell[c]{0.471 \footnotesize(0.019)}
& \makecell[c]{0.591 \footnotesize(0.033)}\\
\hline

\multirow{5}{*}{\makecell[c]{3D Graph}}
& SchNet \cite{schnet}
& \makecell{0.667 \footnotesize(0.021)}
& \makecell{0.881 \footnotesize(0.026)}
& \makecell{0.834 \footnotesize(0.012)}
& \makecell{0.765 \footnotesize(0.034)}
& \makecell{0.467 \footnotesize(0.025)}
& \makecell{0.527 \footnotesize(0.062)}
& \makecell{0.456 \footnotesize(0.024)}
& \makecell{0.573 \footnotesize(0.043)}\\

& DimeNet \cite{dimenetpp}
& N/R
& \makecell{0.878 \footnotesize(0.025)}
& N/R
& \makecell{0.779 \footnotesize(0.019)}
& N/A
& \makecell{0.541 \footnotesize(0.045)}
& \makecell{0.352 \footnotesize(0.101)}
& N/A\\

& PhysChem \cite{physchem}
& \makecell{0.694 \footnotesize(0.024)}
& \makecell{0.848 \footnotesize(0.032)}
& N/A
& \makecell{0.814 \footnotesize(0.017)}
& N/A
& \makecell{0.511 \footnotesize(0.053)}
& N/A
& N/A\\

& M3GNet \cite{m3gnet}
& N/A
& \makecell{0.857 \footnotesize(0.025)}
& N/A
& \makecell{0.697 \footnotesize(0.029)}
& N/A
& \makecell{0.531 \footnotesize(0.034)}
& N/A
& N/A\\

& FAENet \cite{faenet}
& \makecell{0.670 \footnotesize(0.036)}
& \makecell{0.869 \footnotesize(0.013)}
& \makecell{0.788 \footnotesize(0.020)}
& \makecell{0.708 \footnotesize(0.015)}
& \makecell{0.528 \footnotesize(0.094)}
& \makecell{0.474 \footnotesize(0.020)}
& \makecell{0.437 \footnotesize(0.025)}
& \makecell{0.528 \footnotesize(0.035)}\\
\hline

\multirow{10}{*}{\makecell[c]{2D Graph}}
& GATv2 \cite{gatv2}
& \makecell[c]{0.677 \footnotesize(0.053)}
& \makecell[c]{0.891 \footnotesize(0.020)}
& \makecell[c]{0.828 \footnotesize(0.014)}
& \makecell[c]{0.795 \footnotesize(0.013)}
& \makecell[c]{0.502 \footnotesize(0.063)}
& \makecell[c]{0.498 \footnotesize(0.030)}
& \makecell[c]{0.424 \footnotesize(0.027)}
& \makecell[c]{0.560 \footnotesize(0.036)}\\

& GIN \cite{gin}
& \makecell[c]{0.702 \footnotesize(0.031)}
& \makecell[c]{0.897 \footnotesize(0.022)}
& \makecell[c]{0.833 \footnotesize(0.017)}
& \makecell[c]{0.799 \footnotesize(0.021)}
& \makecell[c]{0.543 \footnotesize(0.080)}
& \makecell[c]{0.515 \footnotesize(0.044)}
& \makecell[c]{0.443 \footnotesize(0.027)}
& \makecell[c]{0.568 \footnotesize(0.020)}\\

& EGC \cite{egc}
& \makecell[c]{0.708 \footnotesize(0.043)}
& \makecell[c]{0.896 \footnotesize(0.017)}
& \makecell[c]{0.838 \footnotesize(0.012)}
& \makecell[c]{0.808 \footnotesize(0.029)}
& \makecell[c]{0.575 \footnotesize(0.045)}
& \makecell[c]{0.497 \footnotesize(0.034)}
& \makecell[c]{0.441 \footnotesize(0.023)}
& \makecell[c]{0.566 \footnotesize(0.017)}\\

& MPNN \cite{eccnn}
& \makecell[c]{0.711 \footnotesize(0.022)}
& \makecell[c]{0.894 \footnotesize(0.023)}
& \makecell[c]{0.830 \footnotesize(0.014)}
& \makecell[c]{0.797 \footnotesize(0.018)}
& \makecell[c]{0.532 \footnotesize(0.064)}
& \makecell[c]{0.469 \footnotesize(0.040)}
& \makecell[c]{0.449 \footnotesize(0.057)}
& \makecell[c]{0.564 \footnotesize(0.031)}\\

& UniMP \cite{unimp}
& \makecell[c]{0.702 \footnotesize(0.030)}
& \makecell[c]{0.886 \footnotesize(0.025)}
& \makecell[c]{0.833 \footnotesize(0.014)}
& \makecell[c]{0.793 \footnotesize(0.027)}
& \makecell[c]{0.504 \footnotesize(0.031)}
& \makecell[c]{0.470 \footnotesize(0.025)}
& \makecell[c]{0.422 \footnotesize(0.061)}
& \makecell[c]{0.579 \footnotesize(0.036)}\\

& FiLM \cite{film}
& \makecell[c]{0.703 \footnotesize(0.048)}
& \makecell[c]{0.894 \footnotesize(0.031)}
& \makecell[c]{0.836 \footnotesize(0.014)}
& \makecell[c]{0.783 \footnotesize(0.046)}
& \makecell[c]{0.526 \footnotesize(0.042)}
& \makecell[c]{0.475 \footnotesize(0.032)}
& \makecell[c]{0.421 \footnotesize(0.050)}
& \makecell[c]{0.568 \footnotesize(0.032)}\\

& MEGNet \cite{megnet}
& \makecell[c]{0.604 \footnotesize(0.023)}
& \makecell[c]{0.889 \footnotesize(0.027)}
& \makecell[c]{0.826 \footnotesize(0.038)}
& \makecell[c]{0.754 \footnotesize(0.026)}
& \makecell[c]{0.574 \footnotesize(0.122)}
& \makecell[c]{0.505 \footnotesize(0.027)}
& \makecell[c]{0.422 \footnotesize(0.032)}
& \makecell[c]{0.607 \footnotesize(0.041)}\\

& DMPNN \cite{directed_mpnn}
& \makecell{0.716 \footnotesize(0.037)}
& \makecell{0.879 \footnotesize(0.013)}
& \makecell{0.820 \footnotesize(0.018)}
& \makecell{0.787 \footnotesize(0.008)}
& \makecell{0.566 \footnotesize(0.098)}
& \makecell{0.521 \footnotesize(0.011)}
& \makecell{0.494 \footnotesize(0.011)}
& 0.605 \footnotesize(0.043)\\

& AttFP \cite{attfp}
& \makecell{0.710 \footnotesize(0.021)}
& \bfseries\makecell{0.909 \footnotesize(0.018)}
& \makecell{0.841 \footnotesize(0.017)}
& \makecell{0.807 \footnotesize(0.013)}
& \bfseries\makecell{0.642 \footnotesize(0.079)}
& \makecell{0.513 \footnotesize(0.016)}
& \makecell{0.456 \footnotesize(0.031)}
& 0.588 \footnotesize(0.032)\\

& HEDMoL
& \bfseries\makecell[c]{0.759 \footnotesize(0.043)}
& \bfseries\makecell[c]{0.914 \footnotesize(0.016)}
& \bfseries\makecell[c]{0.865 \footnotesize(0.014)}
& \bfseries\makecell[c]{0.840 \footnotesize(0.010)}
& \bfseries\makecell[c]{0.663 \footnotesize(0.053)}
& \bfseries\makecell[c]{0.572 \footnotesize(0.035)}
& \bfseries\makecell[c]{0.551 \footnotesize(0.008)}
& \bfseries\makecell[c]{0.639 \footnotesize(0.035)}\\
\hline
\end{tabular}
}
\label{tb:results_R2}
\end{table*}

\textbf{Implementations.}
We converted a molecular structure into an attributed graph $G = (\mathcal{V}, \mathcal{U}, \textbf{X}, \textbf{E})$, where $\mathcal{V}$ is the set of nodes (i.e., atoms), $\mathcal{U}$ is the set of edges (i.e., chemical bonds), $\textbf{X} \in \mathbb{R}^{|\mathcal{V}| \times d}$ is a $d$-dimensional node-feature matrix, and $\textbf{E} \in \mathbb{R}^{|\mathcal{U}| \times l}$ is an $l$-dimensional edge-feature matrix. We used the pre-defined 200-dimensional atomic embeddings \cite{matscholar_embs} with atomic features determined by local molecular environments to define the input node-feature matrix $\textbf{X}$. To assign the edge features, we followed the popular implementation that defines the edge features between the atoms as the 22-dimensional one-hot encoding of the bonding types provided in well-known RDKit library\footnote{https://www.rdkit.org}. We used the junction tree algorithm provided by PyTorch Geometric\footnote{https://pytorch-geometric.readthedocs.io} for the implementation of the substructure decomposition step in HEDMoL.

\begin{table}[ht!]
\centering
\caption{Hyperparameter settings of HEDMoL for each benchmark molecular dataset. $B$ and $\gamma$ are the batch size and weight regularization coefficient of the stochastic gradient descent optimizer. $l_a$ and $l_e$ mean the dimensionality of the graph embeddings for the atom-level molecular graph $G_a$ and the electron-derived substructure graph $G_e$.}
\vspace{-2ex}
\label{tb:hparam_settings}
\small
\begin{tabular}{c|cccccccc}
\hline
Dataset
& $\psi_a$
& $\psi_e$
& \makecell{B}
& \makecell{$\gamma$}
& $l_a$
& $l_e$
& $\alpha$
& $\lambda$\\
\hline

Lipop
& EGC
& GIN
& 64
& 5e-6
& 32
& 32
& 0.2
& 1.0\\

ESOL
& EGC
& GIN
& 64
& 5e-6
& 16
& 16
& 0.2
& 1.0\\

ADMET
& EGC
& GIN
& 64
& 5e-6
& 16
& 16
& 0.2
& 1.0\\

IGC50
& EGC
& GIN
& 32
& 5e-6
& 16
& 16
& 0.2
& 0.4\\

LC50
& SchNet
& GIN
& 32
& 5e-6
& 16
& 16
& 0.1
& 0.2\\

LD50
& SchNet
& GIN
& 64
& 1e-6
& 16
& 16
& 0.2
& 1.0\\

LMC-H
& SchNet
& GIN
& 128
& 5e-6
& 16
& 16
& 0.2
& 1.0\\

LMC-R
& SchNet
& GIN
& 32
& 5e-6
& 32
& 32
& 0.2
& 1.0\\\hline
\end{tabular}
\label{tb:hyperparam_settings}
\end{table}

We used the grid search to optimize the training hyperparameters of the competitor GNNs and HEDMoL, such as batch size and weight regularization coefficient. The initial learning rate in the training of HEDMoL was fixed to 5e-4 for all datasets. In the implementation of HEDMoL, we used GIN as the GNN-based embedding network for the electron-derived substructure graphs for all datasets. However, we used EGC or SchNet as the GNN-based embedding network for the atom-level molecular graph. All competitor GNNs were constructed by stacking one dense layer for node-feature embedding, two node aggregation layers with layer normalization \cite{layer_norm}, and two dense layers for prediction. The number of hidden channels was fixed to 256 for all competitor GNNs. HEDMoL used GIN, EGC, and SchNet with the same architecture of the competitor GNNs, but the number of hidden channels between the node aggregation and dense layers was fixed to 128 for fair comparisons on a similar number of model parameters. The hyperparameter settings of HEDMoL for each dataset are presented in Table \ref{tb:hyperparam_settings}. All experiments were conducted in a machine with Intel i9-12900K CPU, 128G memory, and NVIDIA GeForce RTX 3090 Ti.

\begin{figure*}
    \centering
    \includegraphics[width=0.95\textwidth]{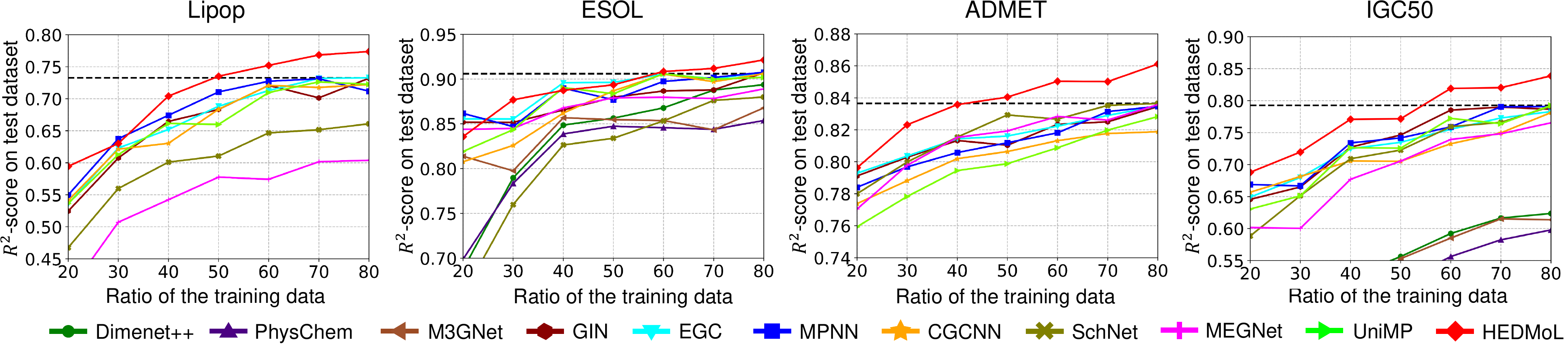}
    \vspace{-2ex}
    \caption{$R^2$-scores for different numbers of the training data. The black dotted line indicates the $R^2$-score of the best competitor method on the 80\% training data.}
    \Description{Prediction accuracy of the competitor models and HEDMoL for different numbers of the training data.}
    \label{fig:R2_small}
\end{figure*}

\subsection{Prediction on Experimentally Collected Benchmark Datasets} \label{sec:exp_benchmark}
We measured $R^2$-scores of the state-of-the-art competitors and HEDMoL based on the 5-fold cross-validation so that the test dataset covers all molecules in the original dataset because the prediction accuracy of the prediction models on the real-world chemical data is sensitive to training/test split \cite{chem_cross_val}. We reported the mean of the measured $R^2$-scores with the standard deviation on the test datasets. Table \ref{tb:results_R2} presents the measured $R^2$-scores and standard deviations on the eight benchmark molecular datasets, and HEDMoL achieved the highest $R^2$-scores for all benchmark datasets. HEDMoL significantly outperformed the competitor methods on the Lipop, ADMET, IGC50, LC50, LD50, LMC-H, and LMC-R datasets. In particular, HEDMoL showed higher $R^2$-scores than individual EGC and GIN, which are used as the embedding networks of HEDMoL. Experimental results on MAEs were consistent with Table \ref{tb:results_R2}, as shown in Appendix \ref{sec:eval_mae}. The experimental results show that HEDMoL learned more generalized and informative molecular representations beyond individual GNNs through the hierarchical representation learning on $G_a$ and $G_e$.

In the experiment, GNNs outperformed the XGB-based models in the problems of predicting the physicochemistry properties on the Lipop, ESOL, and ADMET datasets. However, the simple XGB-based models showed higher $R^2$-scores than those of the state-of-the-art GNNs on the LMC-H and LMC-R datasets, which contain relatively large molecules. This result is consistent with an experimental observation in an existing study \cite{limit_gnn} regarding the overfitting problems of GNNs on large atomic systems. As demonstrated in this experiment, the fingerprint-based models and GNNs have their own limitations when applied to molecular data. The fingerprint-based methods suffer from the lack of physical information about the input molecules because the molecular fingerprints are designed to describe the connectivities of the atoms rather than representing the physical attributes of each atom. By contrast, although GNNs can extract physical information from the input molecular graphs containing physical attributes of each atom, they can be easily overfitted in large atomic systems \cite{limit_gnn}. However, HEDMoL overcomes both limitations by exploiting the molecular graph with electronic attributes, which are robust to extrapolation \cite{dft,dft_mol,challenges_dft}. These results demonstrate that HEDMoL can provide more accurate and robust prediction results on real-world molecular physics without expensive quantum mechanical calculations.

\subsection{Prediction on Small Training Datasets} \label{sec:exp_small}
Since conducting chemical experiments to obtain the experimentally labeled data is time-consuming and labor-intensive, the lack of training data is one of the main challenges of machine learning in chemical applications \cite{tl_chem1,tl_chem2}. Although various methods have been proposed to overcome the lack of training data in chemical applications, existing methods still require manually designed domain knowledge \cite{small_with_knowl}, rigid assumptions on data distributions \cite{small_with_assump}, or DFT calculations \cite{small_with_calc}. On the other hand, as described in the knowledge extension step of Section \ref{sec:knowl_ext}, HEDMoL inherently has the ability to extend the electron-level knowledge regarding small molecules to unseen large molecules, which is beneficial in constructing an accurate prediction model on small training datasets. In this experiment, we compared $R^2$-scores of the competitor methods and HEDMoL for different sizes of training datasets to evaluate the prediction capabilities of HEDMoL on small training datasets.

Fig. \ref{fig:R2_small} shows the $R^2$-scores for different sizes of training datasets. We measured the $R^2$-scores on the Lipop, ESOL, ADMET, and IGC50 datasets in which HEDMoL and most competitor GNNs achieved the $R^2$-scores greater than 0.6. We did not measure the $R^2$-scores of the XGB-based models because they failed on small training datasets. Obviously, we observed that the prediction accuracy tends to be improved as the size of the training dataset increases. However, HEDMoL showed higher $R^2$-scores than the competitor GNNs for all sizes of the training datasets on the Lipop, ADMET, and IGC50 datasets. Furthermore, HEDMoL already achieved comparable $R^2$-scores with each best competitor method on the 80\% training data (black dotted line) on the 50\%, 40\%, and 60\% training data of the Lipop, ADMET, and IGC50 datasets, respectively. These results on the small training datasets show the practical potential of HEDMoL in real-world chemical applications, which usually suffer from the lack of training data.

\begin{table}[t]
\centering
\caption{Test $R^2$-scores of HEDMoL for the external datasets of different molecular scales. $C$ indicates the maximum number of atoms of the molecules in the subset of the QM9 dataset. The knowledge extension of HEDMoL was implemented with the QM9 subset of $C=6$.}
\vspace{-2ex}
\renewcommand{\arraystretch}{1.6}
\fontsize{8pt}{8pt}\selectfont
\resizebox{0.99\linewidth}{!}{
\begin{tabular}{c|cccccc}
\hline
Dataset
& $C=3$
& $C=4$
& $C=5$
& $C=6$
& $C=7$
& $C=8$\\\hline

Lipop
& 0.732 \footnotesize(0.037)
& 0.723 \footnotesize(0.053)
& 0.735 \footnotesize(0.047)
& 0.736 \footnotesize(0.030)
& 0.738 \footnotesize(0.040)
& 0.736 \footnotesize(0.037)\\

ESOL
& 0.914 \footnotesize(0.018)
& 0.911 \footnotesize(0.016)
& 0.915 \footnotesize(0.015)
& 0.915 \footnotesize(0.016)
& 0.916 \footnotesize(0.015)
& 0.914 \footnotesize(0.019)\\

ADMET
& 0.867 \footnotesize(0.007)
& 0.862 \footnotesize(0.015)
& 0.868 \footnotesize(0.012)
& 0.861 \footnotesize(0.010)
& 0.867 \footnotesize(0.012)
& 0.865 \footnotesize(0.014)\\

IGC50
& 0.829 \footnotesize(0.017)
& 0.833 \footnotesize(0.012)
& 0.828 \footnotesize(0.016)
& 0.835 \footnotesize(0.017)
& 0.825 \footnotesize(0.018)
& 0.831 \footnotesize(0.016)\\
\hline
\end{tabular}}
\label{tb:r2_by_scale}
\end{table}

\subsection{Representation Capabilities for Different Coverages of External Calculation Databases}
As described in Section \ref{sec:knowl_ext}, HEDMoL is basically a representation learning method based on knowledge transfer between the information calculated in different scales. In this experiment, we evaluated the prediction capabilities of HEDMoL for different external calculation databases containing molecules in different scales. To this end, we generated external calculation databases of different molecular scales by selecting the molecules from the QM9 dataset based on the number of atoms. In this experiment, six external calculation databases that contain molecules consisting of three to $C$ atoms were generated, where $C = \{3, 4, 5, 6, 7, 8\}$. Note that the knowledge extension of HEDMoL was implemented with $C=6$. Table \ref{tb:r2_by_scale} shows the measured $R^2$-scores of HEDMoL on the external calculation databases of different molecular scales.

In this experiment, although the QM9 subsets of $C = \{7, 8\}$ contain a larger number of molecules, the $R^2$-scores of HEDMoL were robust to the changes of the external datasets in different molecular scales because the junction tree algorithm in the knowledge extension of HEDMoL can decomposes the input large molecules into tiny substructures. For this reason, most decomposed substructures would have been sufficiently approximated by the small molecules in the QM9 subsets of $C = \{3, 4, 5, 6\}$. Similarly, the changes on the $R^2$-scores were marginal for $C = \{7, 8\}$ because most decomposed substructures will be replaced with the small molecules containing 3-6 atoms even on the subsets containing larger molecules. This experimental result shows that preparing an external dataset for HEDMoL is not challenging because an external dataset containing tiny molecules with 3-6 atoms is sufficient to train HEDMoL.

\begin{table}
\centering
\caption{Ablation studies on HEDMoL. KE: knowledge extension in Section~\ref{sec:knowl_ext}. HRL: hierarchical molecular representation learning in Section~\ref{sec:hedmol_training}. PCR: energy-based physical consistency regularization in Section~\ref{sec:consistency regularization}.}
\vspace{-2ex}
\fontsize{8pt}{8pt}\selectfont
\resizebox{0.99\linewidth}{!}{
\begin{tabular}{c|cc|ccc|c}
\hline
Dataset
& EGC
& GIN
& KE
& KE+HRL
& KE+PCR
& HEDMoL\\\hline

Lipop
& \makecell{0.708\\\footnotesize(0.043)}
& \makecell{0.702\\\footnotesize(0.031)}
& \makecell{0.722\\\footnotesize(0.051)}
& \makecell{0.726\\\footnotesize(0.049)}
& \makecell{0.731\\\footnotesize(0.044)}
& \makecell{0.759\\\footnotesize(0.043)}\\[0.15cm]

ESOL
& \makecell{0.896\\\footnotesize(0.017)}
& \makecell{0.897\\\footnotesize(0.022)}
& \makecell{0.913\\\footnotesize(0.015)}
& \makecell{0.913\\\footnotesize(0.015)}
& \makecell{0.915\\\footnotesize(0.017)}
& \makecell{0.914\\\footnotesize(0.016)}\\[0.15cm]

ADMET
& \makecell{0.838\\\footnotesize(0.012)}
& \makecell{0.833\\\footnotesize(0.017)}
& \makecell{0.862\\\footnotesize(0.014)}
& \makecell{0.864\\\footnotesize(0.015)}
& \makecell{0.860\\\footnotesize(0.013)}
& \makecell{0.865\\\footnotesize(0.014)}\\[0.15cm]

IGC50
& \makecell{0.808\\\footnotesize(0.029)}
& \makecell{0.799\\\footnotesize(0.021)}
& \makecell{0.827\\\footnotesize(0.018)}
& \makecell{0.833\\\footnotesize(0.017)}
& \makecell{0.826\\\footnotesize(0.014)}
& \makecell{0.840\\\footnotesize(0.010)}\\[0.15cm]

LC50
& \makecell{0.575\\\footnotesize(0.045)}
& \makecell{0.543\\\footnotesize(0.080)}
& \makecell{0.631\\\footnotesize(0.091)}
& \makecell{0.640\\\footnotesize(0.084)}
& \makecell{0.603\\\footnotesize(0.061)}
& \makecell{0.663\\\footnotesize(0.053)}\\
\hline
\end{tabular}}
\label{tb:ablation_study}
\vspace{-2ex}
\end{table}

\begin{figure*}
    \centering
    \includegraphics[width=0.85\textwidth]{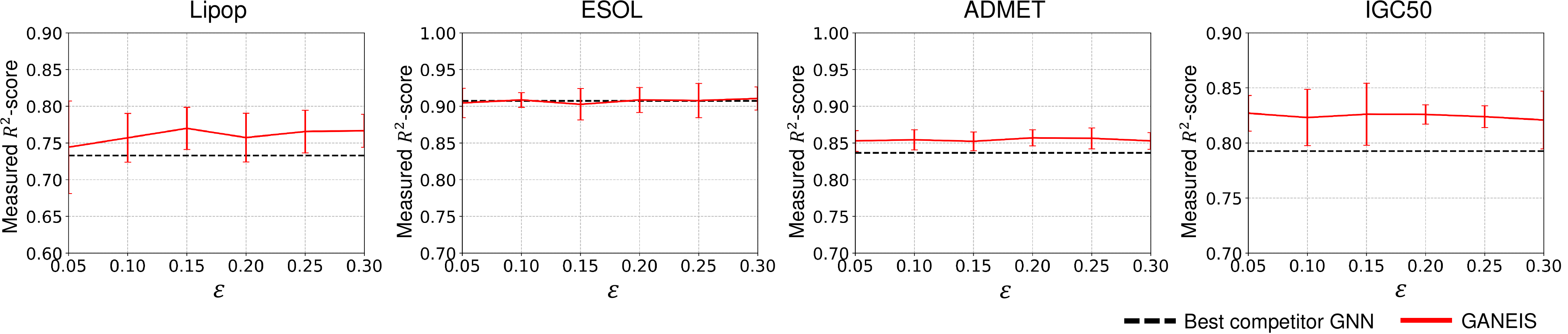}
    \vspace{-2ex}
    \caption{$R^2$-scores of HEDMoL on the Lipop, ESOL, ADMET, and IGC50 datasets for different values of $\alpha$}
    \Description{Prediction accuracy of HEDMoL for different values of $\epsilon$.}
    \label{fig:r2_by_alpha}
\end{figure*}
\begin{figure*}
    \centering
    \includegraphics[width=0.85\textwidth]{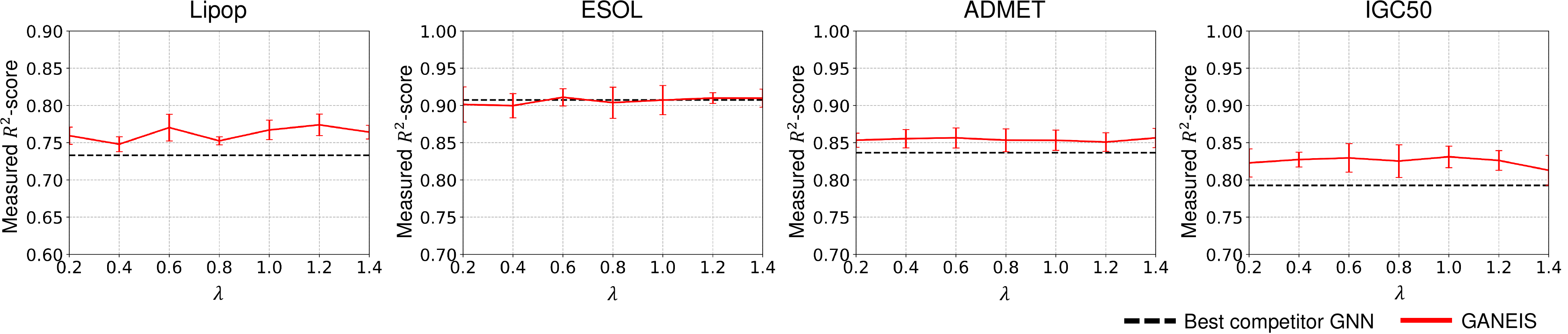}
    \vspace{-2ex}
    \caption{$R^2$-scores of HEDMoL on the Lipop, ESOL, ADMET, and IGC50 datasets for different values of $\lambda$}
    \Description{Prediction accuracy of HEDMoL for different values of $\lambda$.}
    \label{fig:r2_by_lambda}
\end{figure*}

\begin{table*}
\centering
\caption{$R^2$-scores of HEDMoL for different implementations of the knowledge extension. FEATURE-G, FGSD, Graph2Vec, and GeoScattering are unsupervised graph embedding methods.}
\renewcommand{\arraystretch}{1.6}
\fontsize{8pt}{8pt}\selectfont
\begin{tabular}{c|ccccc}
\hline
Dataset
& Morgan \cite{ecfp}
& FEATHER-G \cite{feat_g}
& FGSD \cite{fgsd}
& Graph2Vec \cite{graph2vec}
& GeoScattering \cite{geos}\\\hline

Lipop
& 0.734 \footnotesize(0.044)
& 0.734 \footnotesize(0.042)
& 0.741 \footnotesize(0.038)
& 0.734 \footnotesize(0.047)
& 0.736 \footnotesize(0.030)\\

ESOL
& 0.909 \footnotesize(0.018)
& 0.912 \footnotesize(0.019)
& 0.914 \footnotesize(0.014)
& 0.910 \footnotesize(0.016)
& 0.915 \footnotesize(0.016)\\

ADMET
& 0.865 \footnotesize(0.011)
& 0.870 \footnotesize(0.010)
& 0.864 \footnotesize(0.013)
& 0.864 \footnotesize(0.010)
& 0.861 \footnotesize(0.010)\\

IGC50
& 0.796 \footnotesize(0.018)
& 0.835 \footnotesize(0.011)
& 0.837 \footnotesize(0.021)
& 0.831 \footnotesize(0.023)
& 0.835 \footnotesize(0.017)\\
\hline
\end{tabular}
\label{tb:r2_by_emb}
\end{table*}

\subsection{Ablation Study}
In this experiment, we conducted an ablation study of HEDMoL to investigate the effects of each component on the prediction capabilities of HEDMoL. We compared the $R^2$-scores of EGC and GIN because HEDMoL is equivalent to the embedding networks $\psi_a$ or $\psi_e$ if all the proposed components are removed from HEDMoL. Table \ref{tb:ablation_study} shows the experimental results on HEDMoL. To implement KE without HRL, we used $\textbf{z} = \textbf{z}_a \oplus \textbf{z}_e$ as the final molecular embedding instead of $\textbf{z} = \textbf{z}_a \oplus \textbf{z}_c$. We observed that the $R^2$-scores of the prediction models notably increased after implementing KE for all datasets, as shown in Table \ref{tb:ablation_study}. This result explicitly shows the effectiveness of our knowledge extension approach for propagating the electron-level attributes to the atom-level molecular structures. Although the improvements by KE+HRL and KE+PCR were marginal in the benchmark datasets, the $R^2$-scores were notably improved by HEDMoL (KE+HRL+PCR) on the Lipop, IGC50, and LC50 datasets.

\subsection{Hyperparameter Analysis} \label{sec:hyperparam_analysis}
In the training process of HEDMoL, there are two hyperparameters $\alpha$ and $\lambda$ to control the structural approximation errors and the effect of the physical consistency regularization, respectively. In this experiment, we measured the $R^2$-scores of HEDMoL for different values of $\alpha$ and $\lambda$ on the Lipop, ADMET, and IGC50 datasets, which contain relatively large number of molecules. We selected the hyperparameter values of $\alpha$ and $\lambda$ in \{0, 0.05, 0.1, 0.15, 0.2, 0.25, 0.3, 0.35\} and \{0.2, 0.4, 0.6, 0.8, 1.0\}, respectively. Figs. \ref{fig:r2_by_alpha} and \ref{fig:r2_by_lambda} show the measured $R^2$-scores for different hyperparameter values. In this experiment, HEDMoL consistently achieved better $R^2$-scores than those of the best competitor GNNs for all hyperparameter values and benchmark datasets. In particular, HEDMoL outperformed the best competitor GNNs over the standard deviations of the $R^2$-scores on the ADMET and IGC50 datasets for all values of $\alpha$ and $\lambda$. These experimental results show the robustness of HEDMoL to the hyperparameter values.

\begin{table*}
\centering
\caption{Execution time per epoch (sec/epoch) on the LC50, Lipop, and LD50 datasets. The LC50, Lipop, and LD50 datasets contain 822, 4,200, 7,412 molecules, respectively.}
\vspace{-2ex}
\renewcommand{\arraystretch}{1.6}
\fontsize{8pt}{8pt}\selectfont
\begin{tabular}{c|ccccc|ccccc}
\hline
\multirow{2}{*}{Dataset}
& \multicolumn{5}{c|}{Training}
& \multicolumn{5}{c}{Inference}\\
\cline{2-11} 

& \multicolumn{1}{c}{EGC}
& \multicolumn{1}{c}{GIN}
& \multicolumn{1}{c}{MPNN}
& \multicolumn{1}{c}{SchNet}
& HEDMoL
& \multicolumn{1}{c}{EGC}
& \multicolumn{1}{c}{GIN}
& \multicolumn{1}{c}{MPNN}
& \multicolumn{1}{c}{SchNet}
& HEDMoL \\
\hline

LC50
& \multicolumn{1}{c}{0.087}
& \multicolumn{1}{c}{0.077}
& \multicolumn{1}{c}{0.091}
& \multicolumn{1}{c}{0.172}
& 0.229
& \multicolumn{1}{c}{0.014}
& \multicolumn{1}{c}{0.013}
& \multicolumn{1}{c}{0.014}
& \multicolumn{1}{c}{0.022}
& 0.035\\

Lipop
& \multicolumn{1}{c}{0.643}
& \multicolumn{1}{c}{0.558}
& \multicolumn{1}{c}{0.619}
& \multicolumn{1}{c}{1.650}
& 1.371
& \multicolumn{1}{c}{0.099}
& \multicolumn{1}{c}{0.090}
& \multicolumn{1}{c}{0.099}
& \multicolumn{1}{c}{0.191}
& 0.228\\

LD50
& \multicolumn{1}{c}{0.929}
& \multicolumn{1}{c}{0.828}
& \multicolumn{1}{c}{0.902}
& \multicolumn{1}{c}{2.038}
& 2.468
& \multicolumn{1}{c}{0.136}
& \multicolumn{1}{c}{0.125}
& \multicolumn{1}{c}{0.133}
& \multicolumn{1}{c}{0.237}
& 0.373\\
\hline
\end{tabular}
\label{tb:time_per_epoch}
\end{table*}

\subsection{Prediction Capabilities for Different Graph Embedding Methods in the Knowledge Extension Step} \label{sec:r2_by_emb}
There are several choices of $\pi (S_k, G_i^s)$ in the implementation of the knowledge extension in HEDMoL. In this experiment, we evaluated the prediction capabilities of HEDMoL for different graph embedding methods in the knowledge extension step. Additionally, we evaluated an implementation of $\pi (S_k, G_i^s)$ based on the Tanimoto distance \cite{tanimoto_sim} for the Morgan fingerprints \cite{ecfp} of the molecules, which is a common approach for finding similar molecules in cheminformatics. This experimental evaluation was conducted on the Lipop, ESOL, ADMET, and IGC50 datasets, where the competitor GNNs and HEDMoL achieved R2 scores greater than 0.7. As shown in Table \ref{tb:r2_by_emb}, HEDMoL with the graph embedding methods showed similar $R^2$-scores for all benchmark datasets because HEDMoL already assumes the matching errors in the knowledge extension, which was formalized by the Gaussian error $\epsilon_k$. As a result, HEDMoL was robust to the implementation of the knowledge extension.

\subsection{Execution Time for Training/Inference}
HEDMoL inevitably requires additional computation costs in the training and inference phases to execute the hierarchical representation learning with two embedding networks $\psi_a$ and $\psi_e$. Moreover, calculating the regularization terms of the physical consistency also causes additional computations. We compared the execution time of HEDMoL with EGC, GIN, MPNN, and SchNet on the LC50, Lipop, and LD50 datasets. MPNN and SchNet showed the best prediction accuracy on the benchmark datasets. EGC, GIN, and SchNet were used as the embedding networks of HEDMoL. The LC50, Lipop, and LD50 datasets contain 822, 4,200, and 7,412 molecules, respectively. This experiment was conducted in a machine with Intel i9-12900K CPU, 128G memory, and NVIDIA GeForce RTX 3090 Ti.

Table \ref{tb:time_per_epoch} shows the execution time of MPNN, SchNet, and HEDMoL in the training and inference processes. HEDMoL requires more execution time in the training and inference processes because it has two GNN-based embedding networks. Nonetheless, the execution time of HEDMoL linearly increased. For example, the execution times of HEDMoL on the Lipop dataset was similar to the sum of the execution times of EGC and GIN, as shown in $1.371 \approx 0.643+0.558$ and $0.228 \approx 0.099+0.090$. These experimental results show that the forward process of HEDMoL is as efficient as the forward processes of existing GNNs.

\section{Conclusion}
This paper proposed HEDMoL for learning electron-derived molecular representations to improve the prediction capabilities on real-world molecular physics. HEDMoL learned the electron-derived molecular representations without additional calculation costs by 
transferring the pre-calculated electron-level information of small molecules in an external database to large input molecules. Based on the electron-derived molecular representations, HEDMoL achieved state-of-the-art prediction accuracy on extensive molecular datasets containing experimentally observed molecules and their properties. Furthermore, HEDMoL achieved better prediction accuracies even under the lack of training data, which is one of the main challenges of machine learning in chemistry. These results showed the practical potential of HEDMoL in real-world chemical applications.

\section*{Acknowledgements}
This research was supported by Korea Research Institute of Chemical Technology (No. KK2451-10). This work was supported by the National Research Foundation of Korea (NRF) grant funded by the Korea government (MSIT) (RS-2024-00335098).

\bibliographystyle{ACM-Reference-Format}
\bibliography{ref}

\onecolumn
\appendix

\section{Competitor Methods} \label{sec:competitor_methods}
In the experiments, we compared the prediction capabilities of HEDMoL with a baseline tree method and eight state-of-the-art GNNs, which have been widely used in chemical applications. The competitor methods are briefly described as:
\begin{itemize}
    \item \textbf{XGB-Mor}: XGBoost (XGB) \cite{xgb} is a tree-based gradient boosting model, and it showed state-of-the-art performances in various scientific applications. For the experimental evaluations, we generated XGB-Mor that predicts the target molecular properties for the Morgan (Mor) fingerprints of the atom-level molecular structures.
    \item \textbf{XGB-FC}: We generated XGB-FC by combining XGB with the functional-class (FC) \cite{ecfp} fingerprints of the input molecules. The FC fingerprint represents the atom-level molecular structures based on their functional substructures and atoms.
    \item \textbf{XGB-MK}: We also generated XGB-MK based on the MACCS Key (MK) fingerprint, which is one of the most commonly used molecular representations. MACCS key encodes the atom-level molecular structures based on 166-bits binary patterns.
    \item \textbf{SchNet} \cite{schnet}: It is a convolutional neural network for learning molecular representations based on quantum interactions in molecules. It has been widely used as a baseline model in various chemical applications \cite{schnet}.
    \item \textbf{DimeNet} \cite{dimenetpp}: Dimenet and its variatants employ a directional message passing scheme to learn directional information in physical chemistry. We used an advanced model of the original DimeNet, called DimeNet++.
    \item \textbf{PhysChem} \cite{physchem}: It is a neural architecture that learns molecular representations via fusing the information about the inter-atomic geometry and message passing through chemical bonds.
    \item \textbf{M3GNet} \cite{m3gnet}: Graph neural networks with three-body interactions (M3GNet) is a neural network to learn molecular representations based on the three-body inter-atomic interactions beyond the conventional single atomic attributes and two-body inter-atomic interactions.
    \item \textbf{FAENet} \cite{faenet}: Frame averaging equivariant GNN (FAENet) is simple and fast GNN optimized for stochastic frame-averaging. FAENet can learn molecular representations by processing atom-relative positions with full flexibility without symmetry-preserving requirements.
    \item \textbf{GATv2} \cite{gatv2}: Graph attention network (GAT) was proposed to learn node and graph embeddings based on self-attention mechanism between the nodes \cite{gat}. We employed GAT version 2 (GATv2), which outperformed the standard GAT in comprehensive experiments.
    \item \textbf{GIN} \cite{gin}: Graph isomorphism network (GIN) is a framework for learning graph representations based on graph isomorphism test.
    \item \textbf{EGC} \cite{egc}: Efficient graph convolution (EGC) is an isotropic GNN based on adaptive filters and aggregation fusion in the node aggregation phase. 
    EGC outperformed common anisotropic GNNs, such as graph attention networks, on benchmark datasets.
    \item \textbf{MPNN} \cite{eccnn}: Message passing neural network is a unified framework of node and edge convolution methods for learning molecular representations on quantum chemistry.
    \item \textbf{UniMP} \cite{unimp}: Unified message passaging (UniMP) is a transformer-based GNN. UniMP showed state-of-the-art prediction capabilities by incorporating feature and label propagation at both training and inference time based on the transformer architecture.
    \item \textbf{FiLM} \cite{film}: Feature-wise Linear Modulation (FiLM) was devised to effectively propagate information in a graph, allowing feature-wise modulation of the passed information.
    \item \textbf{MEGNet} \cite{megnet}: MatErials Graph Network (MEGNet) was proposed to predict the physical and chemical properties of molecular and crystal structures. It significantly improved the prediction accuracy by propagating the global state of the atomic structures through a message passing process between the atoms.
    \item \textbf{DMPNN} \cite{directed_mpnn}: Directed MPNN (D-MPNN) is an extension of the original MPNN for the directed molecular graphs. DMPNN uses a message passing scheme via directed edges to learn directional information between the atoms.
    \item \textbf{AttFP} \cite{attfp}: AttFP (originally AttentiveFP) employs a graph self-attention mechanism to learn molecular representations in drug-like molecules. AttFP was devised to learn non-local intra-molecular interactions to extract informative molecular representations.
\end{itemize}

\section{Implementation Details} \label{sec:settings}
In the implementation of HEDMoL, we select molecules that consist of six or fewer atoms from the QM9 dataset to obtain a subset containing 685 molecules. This is because we aim to perform knowledge extension based on small molecules since the electron-level calculation error is small for small molecules. In the knowledge extension, the transferred substructure feature $\textbf{X}_{e,k}$ is defined as a substructure-feature matrix $\textbf{X}_e$ is defined as a 12-dimensional vector containing physically calculated electronic attributes: dipole moment, isotropic polarizability, HOMO, MUMO, HOMO-LUMO gap, electronic spatial extent, vibrational energy, internal energy at 0 K and 298.15 K, enthalpy, free energy, and heat capacity.

\section{Prediction Accuracy of HEDMoL Architectures with Junction Tree and BRICS Decomposition Algorithms}
We conducted an experiment with a variant of HEDMoL, i.e., HEDMoL-BC, in which the junction tree algorithm in the substructure decomposition process is replaced with the BRICS decomposition \cite{brics}. The BRICS decomposition is the most common graph decomposition method in chemical science to split the molecular graph into the chemically-valid substructures.

\begin{table}[ht!]
\centering
\caption{Measured R2-scores of HEDMoL and HEDMoL-BC.}
\label{tb:r2_brics}
\renewcommand{\arraystretch}{1.1}
\small
\begin{tabular}{c|cccccccc}
\hline
Method
& Lipop
& ESOL
& ADMET
& IGC50
& LC50
& LD50
& LMC-H
& LMC-R\\
\hline

HEDMoL-BC
& \makecell{0.731 \footnotesize(0.054)}
& \makecell{0.915 \footnotesize(0.021)}
& \makecell{0.862 \footnotesize(0.007)}
& \makecell{0.836 \footnotesize(0.016)}
& \makecell{0.620 \footnotesize(0.081)}
& N/A
& \makecell{0.532 \footnotesize(0.013)}
& \makecell{0.622 \footnotesize(0.039)}\\

HEDMoL
& \bfseries\makecell{0.759 \footnotesize(0.043)}
& \makecell{0.914 \footnotesize(0.016)}
& \makecell{0.865 \footnotesize(0.014)}
& \makecell{0.840 \footnotesize(0.010)}
& \bfseries\makecell{0.663 \footnotesize(0.053)}
& \bfseries\makecell{0.572 \footnotesize(0.035)}
& \makecell{0.551 \footnotesize(0.008)}
& \makecell{0.639 \footnotesize(0.035)}\\
\hline
\end{tabular}
\end{table}

As shown in Table \ref{tb:r2_brics}, HEDMoL showed marginally better prediction accuracy than HEDMoL-BC on the most benchmark datasets. In particular, HEDMoL outperformed HEDMoL-BC on the Lipop and LC50 dataset. This experimental result comes from two reasons: (1) The BRICS decomposition cannot split the large molecules into sufficiently small substructures because it should guarantee the chemical validity of the decomposed substructures. (2) Many hydrogens are virtually added to the decomposed substructures for their chemical validity, and it can distort the structural information of the original molecules. Furthermore, we were not able to execute HEDMoL-BC on the LD50 dataset because the execution time of the BRICS decomposition exploded on several large molecules of the LD50 dataset. This experimental result implies the effectiveness of processing the molecular structures as an attributed graph rather than a chemical system.

\section{Evaluation Results on Mean Absolute Error} \label{sec:eval_mae}
Table \ref{tb:results_mae} presents the measured mean absolute errors (MAEs) of the competitor methods and HEDMoL on the benchmark molecular datasets. As with the result measured by $R^2$-score, HEDMoL showed the best performance for all benchmark datasets, i.e., HEDMoL achieved the lowest MAEs for all benchmark datasets.

\begin{table}[ht!]
\centering
\caption{Measured $R^2$-scores on the benchmark molecular datasets. Input type means the required data format of the input molecules. The highest $R^2$-score for each benchmark dataset has been remarked in bold, and the standard deviation of the $R^2$-scores is presented in parentheses.}
\resizebox{1.0\textwidth}{!}{
\renewcommand{\arraystretch}{1.1}
\begin{tabular}{c|c|ccccccccc}
\hline
Input Type
& Method
& Lipop
& ESOL
& ADMET
& IGC50
& LC50
& LD50
& LMC-H
& LMC-R\\
\hline

\multirow{3}{*}{\makecell[c]{Molecular\\Fingerprint}}
& XGB-Mor \cite{ecfp}
& \makecell[c]{0.608 \footnotesize(0.012)}
& \makecell[c]{0.908 \footnotesize(0.039)}
& \makecell[c]{0.855 \footnotesize(0.018)}
& \makecell[c]{0.648 \footnotesize(0.017)}
& \makecell[c]{0.876 \footnotesize(0.061)}
& \makecell[c]{0.475 \footnotesize(0.008)}
& \makecell[c]{0.358 \footnotesize(0.006)}
& \makecell[c]{0.368 \footnotesize(0.027)}\\

& XGB-FC \cite{ecfp}
& \makecell[c]{0.586 \footnotesize(0.010)}
& \makecell[c]{0.874 \footnotesize(0.064)}
& \makecell[c]{0.855 \footnotesize(0.017)}
& \makecell[c]{0.468 \footnotesize(0.021)}
& \makecell[c]{0.776 \footnotesize(0.020)}
& \makecell[c]{0.484 \footnotesize(0.006)}
& \makecell[c]{0.357 \footnotesize(0.005)}
& \makecell[c]{0.365 \footnotesize(0.007)}\\

& XGB-MK \cite{maccs_key}
& \makecell[c]{0.610 \footnotesize(0.014)}
& \makecell[c]{0.740 \footnotesize(0.047)}
& \makecell[c]{0.773 \footnotesize(0.017)}
& \makecell[c]{0.444 \footnotesize(0.012)}
& \makecell[c]{0.770 \footnotesize(0.139)}
& \makecell[c]{0.491 \footnotesize(0.015)}
& \makecell[c]{0.364 \footnotesize(0.004)}
& \makecell[c]{0.373 \footnotesize(0.018)}\\
\hline

\multirow{5}{*}{\makecell[c]{3D Graph}}
& SchNet \cite{schnet}
& \makecell[c]{0.560 \footnotesize(0.013)}
& \makecell[c]{0.554 \footnotesize(0.042)}
& \makecell[c]{0.606 \footnotesize(0.020)}
& \makecell[c]{0.390 \footnotesize(0.022)}
& \makecell[c]{0.677 \footnotesize(0.019)}
& \makecell[c]{0.483 \footnotesize(0.010)}
& \makecell[c]{0.359 \footnotesize(0.007)}
& \makecell[c]{0.410 \footnotesize(0.016)}\\

& DimeNet \cite{dimenetpp}
& N/A
& \makecell{0.562 \footnotesize(0.045)}
& N/A
& \makecell{0.382 \footnotesize(0.021)}
& N/A
& \makecell{0.468 \footnotesize(0.009)}
& \makecell{0.420 \footnotesize(0.011)}
& N/A\\

& PhysChem \cite{physchem}
& \makecell{0.535 \footnotesize(0.014)}
& \makecell{0.592 \footnotesize(0.041)}
& N/A
& \makecell{0.313 \footnotesize(0.017)}
& N/A
& \makecell{0.478 \footnotesize(0.015)}
& N/A
& N/A\\

& M3GNet \cite{m3gnet}
& N/A
& \makecell{0.585 \footnotesize(0.039)}
& N/A
& \makecell{0.452 \footnotesize(0.021)}
& N/A
& \makecell{0.481 \footnotesize(0.008)}
& N/A
& N/A\\

& FAENet \cite{faenet}
& \makecell{0.558 \footnotesize(0.012)}
& \makecell{0.565 \footnotesize(0.042)}
& \makecell{0.725 \footnotesize(0.018)}
& \makecell{0.443 \footnotesize(0.025)}
& \makecell{0.665 \footnotesize(0.021)}
& \makecell{0.493 \footnotesize(0.009)}
& \makecell{0.378 \footnotesize(0.015)}
& \makecell{0.447 \footnotesize(0.020)}\\
\hline

\multirow{10}{*}{\makecell[c]{2D Graph}}
& GATv2 \cite{gatv2}
& \makecell[c]{0.463 \footnotesize(0.016)}
& \makecell[c]{0.461 \footnotesize(0.021)}
& \makecell[c]{0.609 \footnotesize(0.016)}
& \makecell[c]{0.315 \footnotesize(0.013)}
& \makecell[c]{0.680 \footnotesize(0.024)}
& \makecell[c]{0.484 \footnotesize(0.008)}
& \makecell[c]{0.374 \footnotesize(0.006)}
& \makecell[c]{0.386 \footnotesize(0.009)}\\

& GIN \cite{gin}
& \makecell[c]{0.437 \footnotesize(0.024)}
& \makecell[c]{0.443 \footnotesize(0.040)}
& \makecell[c]{0.575 \footnotesize(0.017)}
& \makecell[c]{0.311 \footnotesize(0.012)}
& \makecell[c]{0.663 \footnotesize(0.038)}
& \makecell[c]{0.475 \footnotesize(0.016)}
& \makecell[c]{0.365 \footnotesize(0.007)}
& \makecell[c]{0.381 \footnotesize(0.008)}\\

& EGC \cite{egc}
& \makecell[c]{0.436 \footnotesize(0.016)}
& \makecell[c]{0.434 \footnotesize(0.016)}
& \makecell[c]{0.590 \footnotesize(0.012)}
& \makecell[c]{0.306 \footnotesize(0.022)}
& \makecell[c]{0.644 \footnotesize(0.039)}
& \makecell[c]{0.486 \footnotesize(0.005)}
& \makecell[c]{0.367 \footnotesize(0.005)}
& \makecell[c]{0.378 \footnotesize(0.009)}\\

& MPNN \cite{eccnn}
& \makecell[c]{0.438 \footnotesize(0.012)}
& \makecell[c]{0.453 \footnotesize(0.017)}
& \makecell[c]{0.611 \footnotesize(0.011)}
& \makecell[c]{0.312 \footnotesize(0.011)}
& \makecell[c]{0.670 \footnotesize(0.053)}
& \makecell[c]{0.500 \footnotesize(0.014)}
& \makecell[c]{0.363 \footnotesize(0.011)}
& \makecell[c]{0.382 \footnotesize(0.014)}\\

& UniMP \cite{unimp}
& \makecell[c]{0.449 \footnotesize(0.020)}
& \makecell[c]{0.472 \footnotesize(0.030)}
& \makecell[c]{0.599 \footnotesize(0.021)}
& \makecell[c]{0.320 \footnotesize(0.024)}
& \makecell[c]{0.682 \footnotesize(0.035)}
& \makecell[c]{0.497 \footnotesize(0.009)}
& \makecell[c]{0.373 \footnotesize(0.011)}
& \makecell[c]{0.376 \footnotesize(0.016)}\\

& FiLM \cite{film}
& \makecell[c]{0.454 \footnotesize(0.018)}
& \makecell[c]{0.450 \footnotesize(0.038)}
& \makecell[c]{0.599 \footnotesize(0.014)}
& \makecell[c]{0.309 \footnotesize(0.023)}
& \makecell[c]{0.695 \footnotesize(0.043)}
& \makecell[c]{0.496 \footnotesize(0.013)}
& \makecell[c]{0.372 \footnotesize(0.014)}
& \makecell[c]{0.383 \footnotesize(0.018)}\\

& MEGNet \cite{megnet}
& \makecell[c]{0.537 \footnotesize(0.019)}
& \makecell[c]{0.474 \footnotesize(0.024)}
& \makecell[c]{0.601 \footnotesize(0.025)}
& \makecell[c]{0.355 \footnotesize(0.018)}
& \makecell[c]{0.662 \footnotesize(0.054)}
& \makecell[c]{0.478 \footnotesize(0.011)}
& \makecell[c]{0.371 \footnotesize(0.008)}
& \makecell[c]{0.366 \footnotesize(0.010)}\\

& DMPNN \cite{directed_mpnn}
& \makecell{0.435 \footnotesize(0.017)}
& \makecell{0.558 \footnotesize(0.038)}
& \makecell{0.595 \footnotesize(0.021)}
& \makecell{0.307 \footnotesize(0.019)}
& \makecell{0.658 \footnotesize(0.043)}
& \makecell{0.475 \footnotesize(0.008)}
& \makecell{0.352 \footnotesize(0.015)}
& 0.365 \footnotesize(0.020)\\

& AttFP \cite{attfp}
& \makecell{0.435 \footnotesize(0.018)}
& \bfseries\makecell{0.435 \footnotesize(0.021)}
& \makecell{0.581 \footnotesize(0.018)}
& \makecell{0.303 \footnotesize(0.022)}
& \bfseries\makecell{0.623 \footnotesize(0.038)}
& \makecell{0.478 \footnotesize(0.016)}
& \makecell{0.357 \footnotesize(0.010)}
& 0.382 \footnotesize(0.018)\\

& HEDMoL
& \bfseries\makecell[c]{0.392 \footnotesize(0.017)}
& \bfseries\makecell[c]{0.427 \footnotesize(0.027)}
& \bfseries\makecell[c]{0.554 \footnotesize(0.026)}
& \bfseries\makecell[c]{0.287 \footnotesize(0.007)}
& \bfseries\makecell[c]{0.603 \footnotesize(0.036)}
& \bfseries\makecell[c]{0.451 \footnotesize(0.013)}
& \bfseries\makecell[c]{0.335 \footnotesize(0.004)}
& \bfseries\makecell[c]{0.344 \footnotesize(0.016)}\\
\hline
\end{tabular}
}
\label{tb:results_mae}
\end{table}
\end{document}